# Catalyst-mediated etching of carbon nanotubes exhibiting electronic-structure insensitivity and reciprocal kinetics with growth

Keigo Otsuka [a,*] and Shigeo Maruyama [a,b,c]

[a] *Department of Mechanical Engineering, The University of Tokyo, Tokyo, 113-8656, Japan*

[b] *School of Mechanical Engineering, Zhejiang University, Hangzhou, 310027, China*

[c] *Institutes of Innovation for Future Society, Nagoya University, Nagoya 464-8601, Japan*

(Dated: 27 April 2025)

The selective etching of carbon nanotubes has been widely explored as a post-synthetic route for enriching semiconducting species. As nanoelectronic applications increasingly demand pure semiconducting nanotubes for use in field-effect transistors and other optoelectronic devices, understanding the mechanistic basis of selective removal becomes critical. While etching selectivity is often attributed to electronic structure effects on tube walls, its relevance in the presence of catalyst nanoparticles remains unclear. Here, we directly quantify the catalyst-mediated etching and growth rates of individual single-walled carbon nanotubes using elaborate isotope labeling methods. Surprisingly, in water vapor and methanol environments, catalytic etching proceeds with negligible dependence on tube electronic type, in sharp contrast to non-catalytic oxidation pathways. *In-situ* Raman analysis upon heating on nanotube ensembles also confirms metallicity-insensitive etching under catalytic conditions, whereas sidewall oxidation without catalysts exhibits pronounced selectivity. Our growth kinetic model, which precisely describes the kinetics of catalytic etching process, motivates kinetic Monte Carlo simulations of nanotube edge dynamics, revealing the reciprocal nature of edge configuration during growth and etching. These findings highlight a fundamental mechanistic distinction between catalytic and non-catalytic reactivity and thus propose that catalytic etching may serve as a diagnostic mirror of growth behavior when using pre-sorted carbon nanotube samples.

1. **Introduction**

Semiconducting single-walled carbon nanotubes (s-SWCNTs) are regarded as a promising candidate for next-generation nanoelectronic devices, particularly in ultrascaled field-effect transistor applications [1]. Their superior transport properties and atomic-scale thinness have inspired extensive research into techniques for enriching semiconducting species from as-grown mixtures [2–10]. Among these, selective chemical etching based on electronic structure—notably the preferential removal of metallic SWCNTs (m-SWCNTs)—has been a powerful route to enhance device performance [11–17]. A substantial body of prior work has reported that m-SWCNTs undergo oxidation more readily than the semiconducting counterparts in the presence of oxidants such as oxygen [14], water ($H_2O$) [16], or carbon dioxide ($CO_2$) [17], enabling easy post-growth purification methods. These observations are typically rationalized in terms of electronic features such as density of states and Fermi level positioning [18,19], all of which may plausibly modulate chemical reactivity at the nanotube surface.

While such interpretations are well supported in the case of sidewall oxidation, it remains less clear whether the similar electronic-structure dependence persists under catalytic conditions [12,13], particularly in the presence of metal nanoparticles and mild oxidants. Since the oxidants do not directly act on the nanotube sidewall with specific electronic structures, reactivity at tube-catalyst and catalyst-molecule interfaces should play a more critical role in the total kinetics of catalytic etching. This regime is relevant not only for selective etching as a post-synthetic treatment, but also for processes governing the growth kinetics of SWCNTs [20]. In many prior studies, interpretations of selectivity in catalytic systems have drawn analogies to non-catalytic oxidation [21–23], often implicitly assuming similar mechanisms without a direct experimental verification.

Insights from crystal growth of silicon may provide useful perspective in this context [24–26]. In catalyst-free epitaxy or solid-phase regrowth, a wealth of experimental evidence has shown that the growth rate is sensitive to doping, and by extension, to shifts in the Fermi level. For instance, compensation doping experiments have clearly demonstrated that the acceleration or suppression of growth kinetics upon substitutional doping can be reversed by restoring the Fermi level to its intrinsic position [24], indicating the dominant role of electronic effects at the growth front. On the other hand, vapor-liquid-solid growth of silicon nanowires, despite often exhibiting growth rate changes upon doping, presents a more nuanced picture. In such systems, it is suggested that the observed kinetic variations frequently stem from changes in dopant segregation, or precursor decomposition behavior [27,28]. In particular, when the rate-limiting step resides in activation on the catalyst surface, rather than at the solid-liquid interface, the impact of Fermi-level positioning in solids on the overall growth rate may become negligible [29].

These observations prompt careful reconsideration of the role of electronic structure in catalytic processes involving other materials, including SWCNTs. Owing to their chirality-dependent band structure, SWCNTs also offer a unique platform where electronic structures can be tuned without introducing extrinsic dopant elements. Although the concept of electronic-type-selective etching has contributed to advances in SWCNT purification and device integration, its applicability under catalytic conditions remains to be critically examined not only for nanotubes, but also across broader catalytic systems.

In this study, we demonstrate that catalyst-mediated etching of SWCNTs proceeds largely independent of electronic type, in sharp contrast to non-catalytic oxidation. To reach this conclusion, we quantify both growth and etching rates of individual SWCNTs with known electronic type, defined as length change per unit time, and find that water vapor environment exhibits no discernible electronic-type dependence under catalytic conditions. Our previously established kinetic model indicates more broadly that growth and etching share symmetric kinetics, and this assumption is supported by kinetic Monte Carlo simulations of edge dynamics, too. We then apply an extended framework to study the dual role of methanol as an etchant and mild carbon source, again finding negligible selectivity. Finally, ensemble-level *in-situ* Raman experiments allow us to directly contrast catalyst-mediated and non-catalytic etching behaviors under identical environments, which reveals that electronic-type selectivity emerges only in the absence of catalysts. Recognizing this distinction is essential for accurately interpreting and designing selective reactions in nanomaterials—for instance, by employing etching experiments as a reciprocal mirror to infer growth process.

## 2. Results and discussion

When SWCNT and nanoparticle are held at high temperature under vacuum, carbon atoms are continuously incorporated into the nanotube lattice or dissolved into the catalyst through reversible interfacial reactions, eventually reaching dynamic equilibrium. If carbon removal into the gas phase is enabled—for example, through reaction with a gaseous oxidant—the equilibrium is disturbed, and the net flux of carbon shifts toward tube shortening as illustrated in Fig. 1a. We define this process as catalytic etching. Among various gas-phase oxidants, water vapor is particularly effective in facilitating carbon removal and has been used in our previous studies [30,31]. Notably, similar reactions of carbon removal into the gas phase implicitly occur during SWCNT growth when using oxygen-containing carbon sources such as ethanol, governing the chirality-dependent growth kinetics [20]. Here, we attempt to isolate the etching process by excluding carbon sources to study the intrinsic kinetics of catalyst-mediated SWCNT etching.

To capture dynamics of growth and etching for identical nanotubes, we employ an isotope labeling technique [32] that enables reconstruction of the length history of individual SWCNTs and identification of their electronic types by Raman mapping measurements. We distinguished s-and m-SWCNTs based on their G-mode lineshape and radial breathing mode (RBM) peak positions (Fig. 1b). Our experimental protocol alternates between controlled growth and etching stages in repeated cycles. As outlined in Fig. 1c, ethanol serves as the carbon source during growth, with $^{13}$C-enriched ethanol supplied at specific intervals to mark the onset of each growth stage. This is followed by an etching stage, during which the ethanol supply is replaced with an etchant (primarily $H_2O$) or an idling stage, during which only Ar/$H_2$ carrier gas flows to maintain a constant total flow rate. Two different levels of $^{13}$C ratios are used to identify the growth after the etching stage and the idling stage. This sequence defines one growth-etching cycle, and we repeat it four times in total.

Time-dependent evolution of nanotube length extracted from these growth-etching cycles is shown in Fig. 1d. A representative s-SWCNT is marked with an arrow in the Raman mapping image of Fig. 1e, whose brightness and color represent the G-mode intensity and its peak frequency, respectively. The corresponding Raman spectra along the nanotube exhibit discrete $^{12}$C-enriched and $^{13}$C-enriched domains, defining the red diamonds and circles in Fig. 1d. We adopt two assumptions for further interpretation of the data obtained. First, the SWCNT growth rate during ethanol supply remains constant before and after the etching stage (blue lines). Second, in the absence of both carbon sources and etchants, the nanotube length is unchanged (gray lines) [30]. Under these constraints, we can clearly conclude that the SWCNTs are shortened due to exposure to water vapor during the etching stage, without any structural changes after the multiple growth-etching transitions (see also Fig. S1).

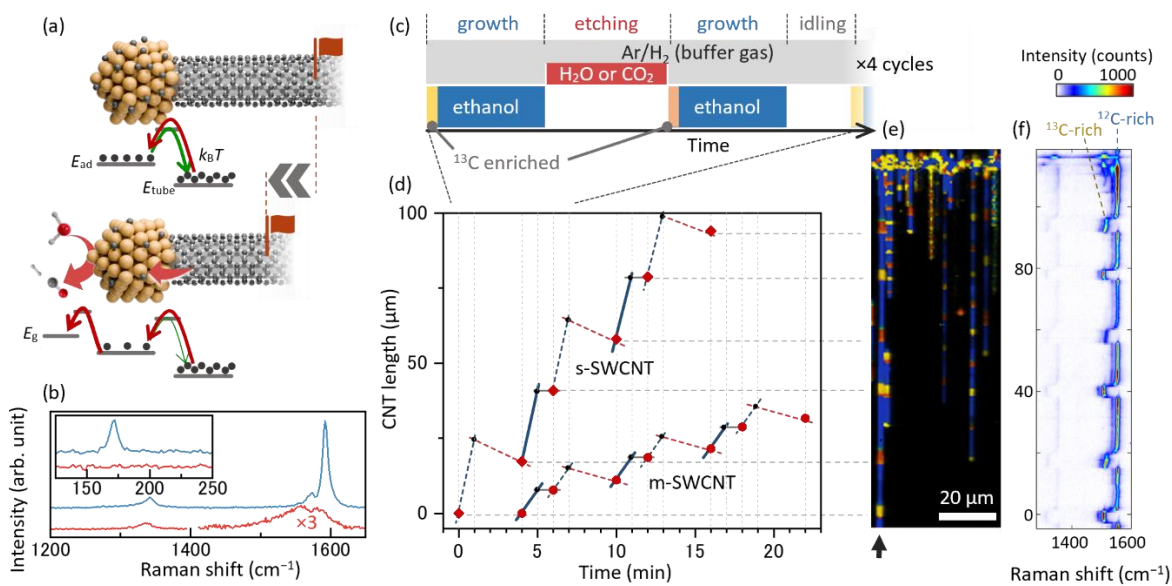

**Fig. 1.** Concept and methodology for probing catalyst-mediated etching of individual SWCNTs. (a) Schematic representation of equilibrium (upper) and etching regimes (lower) between a catalyst and a SWCNT under vacuum and oxidant condition, respectively. Each reaction rate depends on activation barrier and the number of carbon atoms available (black dots). (b) Typical Raman spectra of m- and s-SWCNTs. (c) Sequence of gas supply: alternating periods of ethanol, etchant, and carrier gas (Ar/H$_2$), with periodic use of $^{13}$C-labeled ethanol to trace nanotube growth process. (d) Reconstructed time evolution of nanotube length in the case of water as an etchant. (e) Raman mapping image of a s-SWCNT used for the growth process reconstruction. (f) Spatially resolved Raman spectra showing $^{12}$C- and $^{13}$C-enriched segments along the tube axis.

Using this protocol, we successfully determined both growth and etching rates for a large number of individual SWCNTs. Figure 2a presents histograms of the etching rates for m-SWCNTs (top) and s-SWCNTs (bottom) at the etching stage in water vapor. In both cases, the etching rates follow an approximately normal distribution centered around 1 μm/min. A small number of nanotubes exhibit slightly negative etching rates, that is, apparent growth during the etching stage, which is unreasonable considering the absence of carbon sources. These outliers likely reflect the limitations of our method, arising from temporal fluctuations in growth rate and uncertainties in length determination, rather than nanotube growth in water vapor.

What is particularly notable is the close overlap in etching rate distributions between m- and s-SWCNTs. This observation stands in stark contrast to prior reports on non-catalytic etching by water vapor, in which m-SWCNTs were found to be >10× more reactive than their semiconducting counterparts [15]. The absence of such selectivity here highlights a fundamental difference in the reaction mechanism when catalysis is involved. In addition to water vapor, the etching rate distributions under exposure to $CO_2$ as an etchant also show negligible difference between m- and s-SWCNTs (Fig. S3), reinforcing the conclusion of metallicity-insensitive etching kinetics.

We further examined the relationship between etching and growth rates at the level of individual SWCNTs. Figure 2b shows a scatter plot of these two parameters, revealing a weak but discernible positive correlation. This weak trend likely arises because both growth and etching involve bidirectional carbon exchange between the catalyst and the nanotube lattice, both of which processes share the same rate constant as discussed later. Active area of catalyst nanoparticles is another critical factor connecting these two quantities. The observed deviations from the regression line may reflect variation in rate constants of other steps such as carbon adsorption on catalysts and carbon removal by the oxidant.

To better understand the magnitude of the etching rates, we revisit our previously established kinetic model for SWCNT growth [20], where the net growth rate $\gamma$ is considered as the difference between carbon adsorption rate $\gamma_{ad}$ and desorption rate $\gamma_{de}$ at the catalyst surface. Both $\gamma_{ad}$ and $\gamma_{de}$ are proportional to pressures of involved gas species ($P_C$ and $P_E$, respectively) and their respective rate constants ($k_{ad}$ and $k_{de}$, respectively), and the latter also linearly depends on the carbon concentration on catalyst $N$. From this viewpoint, ethanol has non-zero $k_{ad}$ and $k_{de}$ values. Then, growth and etching can smoothly crossover depending on whether $N$ is higher or lower than the equilibrium level $N_{eq}$, where $\gamma$ is proportional to another rate constant $k_g$, i.e., $\gamma = k_g(N - N_{eq})$. In another prior work [31], we quantified $k_{de}$ of ethanol and water vapor (namely, $k_{de}^{et}$ and $k_{de}^{w}$, respectively) from the changes in growth rates $\gamma$ and found $k_{de}^{w}/k_{de}^{et}$ to be ~1.45. Using this model, we now calculate the expected etching rate as a function of H$_2$O pressure in the absence of ethanol, as shown by the blue curve in Fig. 2b. Interestingly, the experimentally observed etching rates align closely with these predictions, confirming the validity of our previous assumption that $k_g$ is identical during both growth and etching. Although this kinetic symmetry might appear intuitive, it stands in contrast to many conventional crystal growth systems, including a graphene case, where the edge morphologies at growth front differ significantly between growth and etching [33]. In this sense, the nearly symmetric behavior observed in SWCNTs underscores a unique and intriguing feature of a quasi-one-dimensional nanotube system.

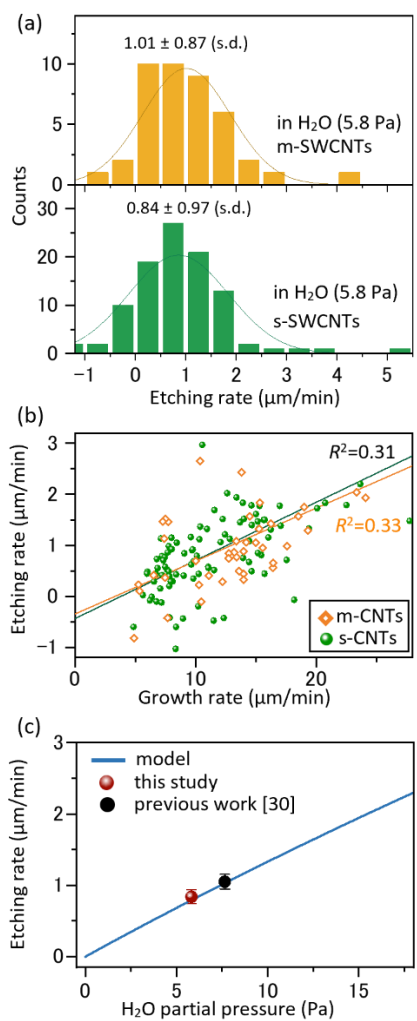

**Fig. 2.** Statistical analysis of etching rates to assess the influence of electronic structure. (a) Histograms of etching rates for m-SWCNTs (upper) and s- SWCNTs (lower) under water vapor, obtained from multiple individual nanotubes. Numbers above the Gaussian fitting lines represent the average and standard deviation of etching rates. (b) Correlation between etching rate and prior growth rate for the same SWCNTs. Positive correlation reflects shared dependence on carbon exchange kinetics at the catalyst-nanotube junction. (c) Average etching rate as a function of water vapor pressure obtained from experiment (circles) and that based on the kinetic model (solid line). Error bars represent standard errors.

To further explore the kinetic symmetry observed between nanotube growth and etching, we conduct kinetic Monte Carlo simulations that explicitly account for the atomic structure of the nanotube edge. Even for a fixed chirality such as (10,10), the open edge can adopt a variety of configurations due to the different sequences of the primitive vectors $\mathbf{a}_1$ and $\mathbf{a}_2$, resulting in distinct arrangements of armchair (AC) and zigzag (ZZ) edge atoms [34–37]. Nanotube growth is primarily driven by carbon dimer ($C_2$) addition at $\mathbf{a}_1$-$\mathbf{a}_2$ anti-AC sites, which enables hexagonal ring closure. In our model, a $C_2$ addition event converts an anti-AC site into an $\mathbf{a}_2$-$\mathbf{a}_1$ AC site, preserving the edge topology while altering its local structure [38]. As shown in Fig. 3b, there are four distinct types of anti-AC motifs depending on their local bonding environment, each associated with an increase (red circles), decrease (blue), or no change (green and yellow) in the total number of (anti-)AC sites. Although these differences could explain chirality-dependent growth rates, a detailed kinetic analysis of edge structures lies beyond the present scope.

We categorized edge configurations by their AC site count $i$, and assigned each group a degeneracy $g_i$, corresponding to the number of distinct configurations [39]. Following prior studies [40], we simply assumed the interfacial energy at a tube-catalyst junction depends only on the numbers of AC and ZZ atoms, neglecting interaction effects of AC-ZZ mixing. $C_2$ addition and removal reactions are simulated using a certain activation energy and an interfacial energy difference $E_Z$–$E_A$ between ZZ and AC edge atoms. Figure 3c shows the time-averaged AC fraction during growth and etching as a function of $E_Z$–$E_A$. The curves overlap precisely, indicating that edge morphology remains statistically unchanged between the two processes. For $E_Z=E_A$, the distribution of AC site counts reproduces the degeneracy count $g_i$ (Fig. 3d), while for nonzero energy differences, the distributions align with a Boltzmann-weighted $g_i$, confirming that the AC fraction is thermodynamically determined by the free energy landscape (Fig. S4).

For comparison, we applied a similar approach to graphene edge. To maintain consistency with the SWCNT model while accommodating the planar symmetry of graphene, we constructed an initial graphene island combining six lattice vectors, which are derived from $\mathbf{a}_1$ by successive 60° rotations, forming a closed hexagonal structure. To avoid kinetic stagnation, known to occur when the edge becomes fully zigzag under $C_2$-only conditions, we included $C_3$ attachment and $C_1$ detachment at ZZ sites, following earlier models [41]. As shown in Fig. 3e, the fraction of AC sites differs significantly between growth and etching, with ZZ edges dominating during growth and AC edges emerging under etching. Representative edge morphologies highlight this asymmetry (Fig. 3f).

These results highlight a unique aspect of nanotubes: the closed-loop edge geometry imposes topological constraints that lead to symmetry in both the kinetics and thermodynamics for growth and etching. This structural feature justifies the assumption of equivalent rate constants $k_g$ for nanotube growth and etching, as supported by our experimental data and kinetic Monte Carlo simulation (see also Fig. S5). These results demonstrate that catalytic etching, far from being a mere post-synthetic treatment, can serve as a dynamic probe for growth kinetics itself, offering a new window into the mechanistic foundation of nanotube synthesis.

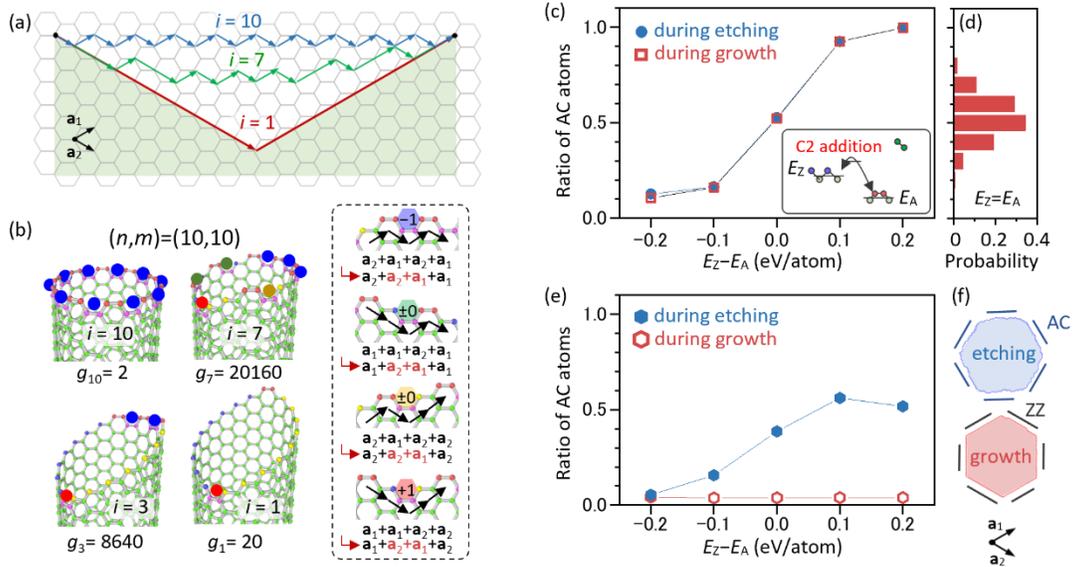

**Fig. 3.** Kinetic Monte Carlo simulations to study the kinetic symmetry for growth and etching. (a) Enumeration of edge configurations for a (10,10) nanotube, based on sequences of $\mathbf{a_1}$ and $\mathbf{a_2}$ primitive vectors. (b) Four representative edge structures of a nanotube. Different types of anti-AC sites for $C_2$ dimer addition are colored in different colors, depending on whether $C_2$ addition change the number of anti-AC sites after the reaction. (c) Kinetic Monte Carlo simulation results for the average fraction of (anti-)AC edge atoms during CNT growth and etching, plotted as a function of edge energy bias ($E_Z$–$E_A$). (d) Distribution of edge structures when $E_Z = E_A$. (e) Corresponding simulation results for graphene. (f) Representative edge configurations for graphene under growth and etching.

Unlike water vapor, alcohols play a dual role [42,43], supplying carbon, while also facilitating its removal, as demonstrated in our prior work (Fig. 4a) [20]. Among various alcohol species, methanol has been widely reported to selectively etch m-SWCNTs [44,45], though its detailed reaction pathways remain uncaptured. This incomplete mechanistic picture prompts a key question: does the reported selectivity reflect a catalytic etching process, a delicate balance between carbon addition and removal at the catalyst surface, or a non-catalytic sidewall oxidation? To address this, we systematically varied the flow rate of methanol while using isotope-labeled ethanol as a reference carbon source, aiming to decouple its supply and removal contributions and assess the effects of electronic structures on nanotube growth kinetics.

As shown in Fig. 4b, we alternately supplied $^{12}$C- and $^{13}$C- enriched ethanol while maintaining a constant total flow rate, and intermittently introduced $^{12}$C-enriched (natural abundance) methanol with different flow rates. The $^{13}$C-based label not only serves to reconstruct nanotube growth history, but also enables quantification of the relative contributions of ethanol and methanol to carbon supply for nanotube lattice, based on G-mode shifts in the resulting Raman spectra.

Figure 4c displays growth curves for three representative SWCNTs reconstructed from spatially resolved Raman mapping images shown in Fig. 4d. In all cases, growth is slowed upon methanol addition. Assuming a constant growth rate during ethanol-only periods, some SWCNTs (e.g., s-CNT #2) exhibit length changes that are best explained by net shortening under methanol exposure, indicating that carbon removal outpaces carbon supply. Interestingly, the G-mode spectra measured within the same nanotube offer complementary but seemingly contradictory insight (Fig. 4e). As methanol flow increases (from segment (2) to (4)), the spectral downshift of G mode due to $^{13}$C from ethanol becomes progressively smaller, suggesting that a noticeable fraction of $^{12}$C from methanol is incorporated into the nanotubes.

We next statistically evaluated methanol's carbon supply capability $k_{ad}$ by comparing Raman shifts under methanol addition to those under ethanol alone. Figure 4f shows that at 3 sccm methanol, corresponding to 60% relative to the ethanol flow, the effective carbon incorporation from methanol accounts for approximately 15–20% of the total carbon, based on the shift amount in G-mode frequency. Comparable results were obtained for m-SWCNTs based on D-mode analysis. Despite this contribution as a carbon source, methanol overall reduces the net growth rate $\gamma$. Figure 4g presents histograms of normalized growth rates (with the ethanol-only rate set to 1) for different flow rates of methanol. No SWCNTs exhibited accelerated growth under methanol addition; some began to go negative at 5 sccm, and the majority entered the etching regime at 7 sccm.

To interpret these results, we again turned to the kinetic model [20]. Figure 4h plots the experimentally determined changes in growth rate $\gamma$ (left axis, orange) and carbon supply contribution $\gamma_{ad}$ (right axis, green) as a function of methanol flow rate $x$. The latter scales linearly with methanol concentration, indicating an average $k_{ad}$ of ~30% compared to ethanol under equivalent partial pressures. We then fit the full set of growth rate data using our kinetic model, incorporating the carbon supply capability as fixed and treating the carbon removal efficiency $\eta = k_{de}^{me}/k_{de}^{et}$, which is the ratio of carbon removal efficiency of methanol and ethanol, as the sole fitting parameter. The dashed curves represent model predictions for various $\eta$ values, with the best fit obtained at $\eta=1.17$, suggesting that methanol exhibits a carbon removal capability comparable to that of ethanol—an outcome that is chemically reasonable, given both molecules possess a single hydroxyl group. Also, examining individual nanotubes can reveal detailed kinetic features that may otherwise be obscured. Figure S3 highlights such a case; s-CNT #2 shows a smooth and monotonic transition from growth to etching with an increasing methanol flow rate, again illustrating the kinetic symmetry between elongation and shortening within a single SWCNT.

Central to the main conclusion of this study, Figure 4i compares the effects of methanol on m- and s-SWCNTs. Neither the carbon supply nor the removal capability of methanol shows any discernible dependence on electronic structure. This result is consistent with the behavior observed in water vapor etching and strongly reinforces the conclusion that iron-mediated reactivity is largely metallicity-insensitive. We note that since the rate constants of methanol-catalyst reactions were determined relative to those of ethanol, if any intrinsic metallicity selectivity in ethanol reactions exists, it would likely be inherited by methanol.

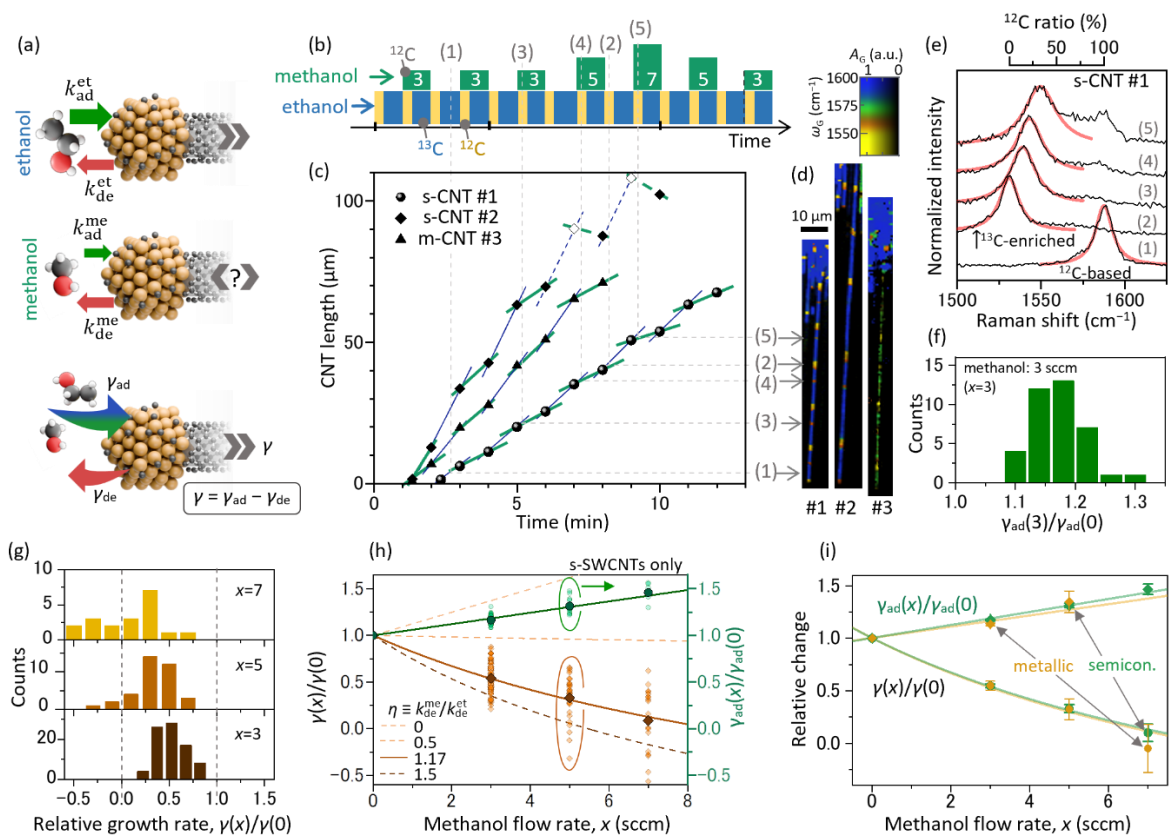

**Fig. 4.** Quantification of the roles of methanol in nanotube growth/etching kinetics. (a) Schematic illustrating methanol's dual role as both carbon source and etchant. (b) SEM image of CNTs. (b) Gas supply protocol alternating $^{12}$C- and $^{13}$C-ethanol, with intermittent methanol addition at varying flow rates. (c) Length profiles over time of three representative SWCNTs reconstructed from isotope labeling. (d) Raman mapping images corresponding to the tubes in (c). (e) Raman

spectra from different segments (1–5) showing decreases in $^{13}$C content in nanotubes with $^{12}$C methanol addition to $^{13}$C ethanol. (f) Distribution of carbon supply rate to catalysts $\gamma_{ad}$ with 3 sccm of methanol addition relative to that with ethanol only. (g) Normalized growth rates under different methanol flow conditions. (h) Relative carbon supply rate (green) and nanotube growth rate (blue) as a function of methanol flow rate. Solid and dashed lines are the model-based curves [20], respectively, representing the influence of $\eta$ on growth kinetics. (i) Comparison of methanol-induced carbon-supply and growth-rate changes between m- and s-SWCNTs, showing negligible differences for both aspects of kinetics.

To further validate our conclusions in a more conventional framework, we investigated the etching behavior of SWCNT networks in water vapor by performing *in-situ* Raman spectroscopy during thermal annealing. In contrast to our earlier analysis on individual tubes, this approach allows statistical evaluation of etching selectivity across a broader ensemble of nanotubes, although decrease in Raman intensity cannot be unambiguously attributed to either a reduction of nanotube number or length, nor can it fully distinguish between catalyst-mediated and non-catalytic etching processes.

We prepared Raman measurement substrates by patterning alignment markers on thermally oxidized silicon substrates. Iron catalysts were deposited in a selected region, while the remainder was left catalyst-free. Separately grown SWCNT networks were then transferred onto this substrate using a polymer-based wet transfer method, ensuring that only the nanotubes, not the original catalysts for growth, were carried over. During *in-situ* Raman measurements in controlled water vapor (~7 Pa), the sample was heated while both catalyst-covered and catalyst-free regions were monitored sequentially at discrete temperature levels. To visualize the thermally induced etching and the roles of catalysts, we extracted the G-mode intensity from each Raman spectrum and plotted it over time in Fig. 5a (lower panel), while the corresponding temperature profile is displayed above. This representation facilitates direct comparison between catalyst-covered and catalyst-free regions. Below 700°C, the G-mode intensity remained largely constant in both regions, though the catalyst-covered area showed a slight reduction. Above 800°C, a marked decrease was observed only in the catalyst-covered region, highlighting the catalytic role in promoting etching at lower temperatures.

We then examined the RBM spectra collected before and after heating (Fig. 5b). While noticeable change in RBM spectra was not observed in the catalyst-free region, the catalyst-covered region exhibited a substantial reduction in overall RBM signal after heating. However, when the spectrum after heating is magnified by a factor of five (thin red line), we find the spectral shape remains nearly unchanged. In particular, the relative intensities of RBM peaks above and below 200 cm$^{-1}$, corresponding to m- and s-SWCNTs, respectively, are preserved. This indicates that, despite significant loss of nanotubes, catalyst-mediated etching proceeds without apparent selectivity toward electronic structure—a somewhat surprising result that aligns with our earlier single-tube observations on kinetics.

To examine whether commonly reported selectivity indeed emerges without catalysts, we increased the water vapor pressure to ~100 Pa and performed stepwise heating (in 25°C increments), recording Raman spectra at room temperature after successive heating steps. As shown in Fig. 5c, signal decay occurred at slightly lower temperatures than at 7 Pa of water. Notably, the RBM peaks above 206 cm$^{-1}$ seem to diminish more rapidly than those below 206 cm$^{-1}$, consistent with previous studies reporting higher reactivity for m-SWCNTs in direct oxidation. To further quantify this trend, we integrated the RBM intensities corresponding to s- and m-SWCNTs and plotted their evolution with each heating step in Fig. 5d. Although both signals decrease upon heating, the intensity ratio of s- to m-SWCNT increased sharply above 700°C, indicating preferential etching of m-SWCNTs under non-catalytic conditions. At the same time, G/D intensity ratio exceeds 150 after non-catalytic etching at 850°C in $H_2O$, indicating the etching preference towards defective SWCNTs in the absence of prominent catalyst effects (Fig. S6).

Although this trend may partially reflect curvature-related reactivity, the emergence of selectivity, which was absent in the catalyst-mediated regime, highlights a fundamental difference in the underlying mechanisms. Interestingly, we find that Cu-mediated etching below 750°C resulted in an increased relative abundance of near-zigzag SWCNTs as shown in Fig. 5e. This is likely because near-armchair species exhibit higher etching rates, similar to the growth process [46], supporting the earlier discussion on the kinetic symmetry at the tube-catalyst interface. The modest increase in the semiconducting-to-metallic intensity ratio (Fig. 5f), which is much less pronounced than that observed under non-catalytic conditions (Fig. 5d), further supports the minor role of tubes' electronic structure in the Cu-mediated etching. To explore the role of other catalysts, similar etching experiments were conducted with seven additional systems, summarized in Figs. S7–S8.

So far, our findings show that catalyst-mediated etching under Fe and $SiO_2$ support proceeds without detectable electronic-type selectivity, this however should not be taken as a universal negation of such reactivity differences. The lack of selectivity likely reflects the specific kinetic regime and reaction environment in this study. Other catalytic systems, including different metals or supports [47], may still engage differently with the intrinsic electronic structure of SWCNTs. For example, when catalysts are unsupported [45,48], the metallicity of SWCNTs could play a critical role in the reactivity at catalyst surfaces by regulating the charge flow from/to the floating catalysts. In parallel, electrostatic gating [49] may offer an orthogonal route to tune nanotube reactivity, potentially allowing direct probing of how transient electronic states affect catalytic responses. Systematic etching studies across diverse catalyst-oxidant combinations, ideally using SWCNTs with pre-sorted chiralities [4,5], may thus serve as a powerful inverse probe of catalytic reactions. Such efforts may not only clarify the role of electronic structure in catalytic systems but also guide the rational design of chirality-selective growth strategies. More broadly, chirality-dependent features of SWCNTs may provide universal insights into the roles of electronic-structure-dependent growth dynamics of other materials.

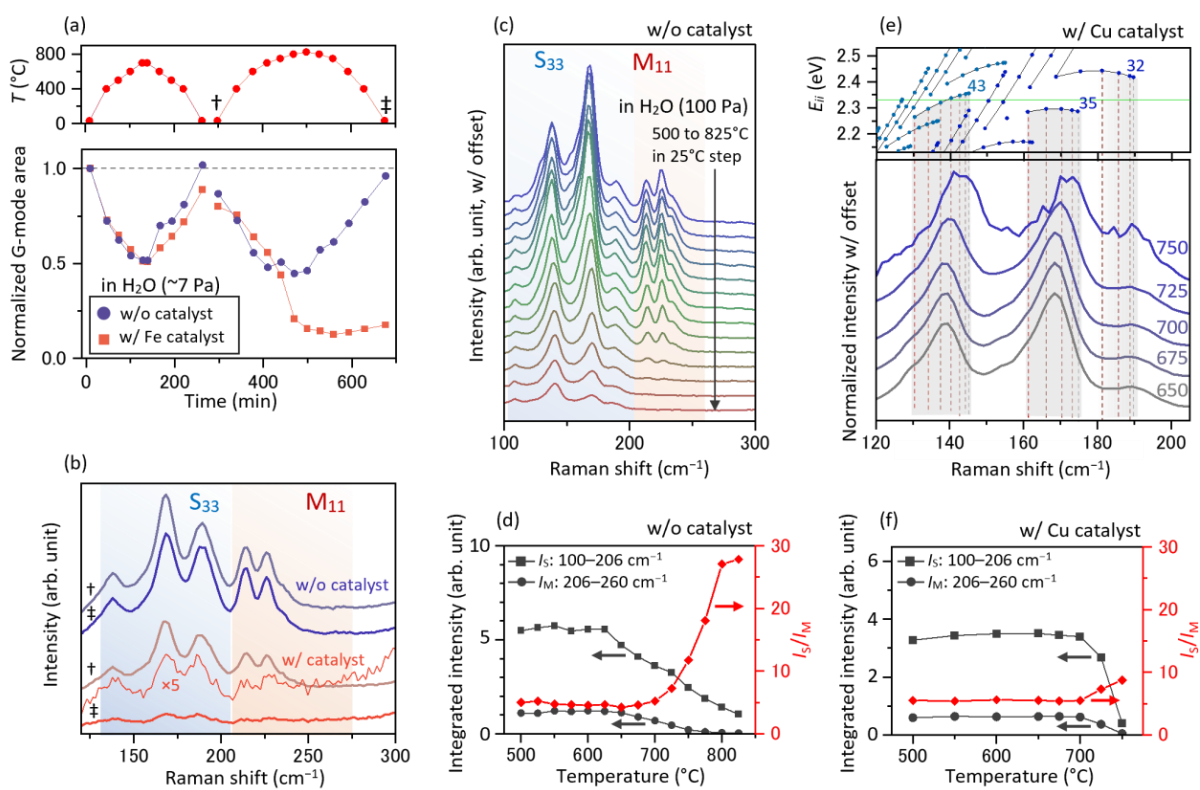

**Fig. 5.** *In-situ* Raman spectroscopy to analyze SWCNT ensemble etching under various catalyst conditions. (a) Time evolution of G-mode intensities measured at both catalyst-covered and catalyst-free regions (lower panel) at different temperatures shown in the upper panel. (b) RBM spectra before and after heating for both regions. The post-heating spectrum for the catalyst-covered region is magnified by a fivefold (thin red line) to highlight spectral features. All spectra are offset for clarity. (c) RBM spectra, where signal from silicon is subtracted, of a catalyst-free region measured after stepwise heating in water vapor at 100 Pa. Each trace was acquired at room temperature following heating steps with a 25°C increment. (d) RBM intensity integrated over the frequency regions corresponding to s-SWCNTs (squares) and m-SWCNTs (circles) measured after heating steps. Red diamonds show the RBM intensity ratio of s-SWCNTs to m-SWCNTs. (e) Normalized Raman spectra obtained after Cu-mediated etching (lower panel) at the indicated temperature (°C). Kataura plot is placed on top, where chiral indices ($n,m$) of the same family type are connected by solid lines. (f) RBM intensity plot similar to (d) in the case of Cu-mediated etching.

## 3. Conclusion

We have examined the role of electronic structure in catalyst-mediated etching of SWCNTs by quantifying growth and etching rates of individual tubes with known electronic types. In contrast to non-catalytic oxidation, water vapor and methanol showed negligible selectivity between s- and m-SWCNTs when etching proceeded via catalyst nanoparticles. However, we emphasize that the observed electronic-type-insensitive reactivity may not be universal, but could be influenced by catalyst composition, oxidant identity, and reaction regime. Our extended growth kinetic model and kinetic Monte Carlo simulations reveal that nanotube elongation and shortening share symmetric kinetics, providing a consistent framework to describe both growth and etching processes. We therefore propose utilizing catalytic etching not merely as a purification technique, but as a diagnostic platform for probing the conditions and selectivity mechanisms relevant to nanotube growth. Broader studies independently employing diverse catalyst systems and pre-sorted SWCNTs could in turn offer new insights for designing strategies towards chirality-selective synthesis.

## 4. Methods

*4.1. Isotope labeling experiments for growth and etching trace.*

Aligned SWCNTs were synthesized on r-cut quartz substrates (Hoffman Materials Inc.) with patterned Fe catalyst stripes (nominal thickness of 0.2 nm, 500 μm pitch) via thermal chemical vapor deposition [32]. While a mixed gas of $Ar/H_2$ (50 sccm, 3% $H_2$) was used as the carrier gas, a carbon precursor were supplied by alternating flows of $^{12}C$ and $^{13}C$ ethanol (Cambridge Isotope Laboratories, 1,2-$^{13}C_2$, 99%) at specific intervals to incorporate isotope labels into the SWCNTs. Growth temperature was maintained at 800°C in a quartz tube. The resultant samples were transferred onto Si substrates with a 100-nm-thick oxide layer using polymer-assisted wet transfer. Prior to the transfer, metal alignment markers were lithographically defined for precise spatial correlation. Raman spectroscopy (Renishaw inVia, 532 nm excitation) was used to reconstruct the growth history of each CNT based on the spectral shifts and the location of each isotope label obtained from mapping measurements [32].

*4.2. In-situ Raman spectroscopy mapping.*

To evaluate catalyst-mediated etching behavior, SWCNT networks were grown separately, then transferred onto patterned Si substrates via polymer-assisted transfer. During the transfer process, iron catalysts were left behind on the original growth substrate, ensuring only SWCNTs were transferred. The receiving substrates were pre-patterned by thermal evaporation using shadow masks into four difference areas: no catalyst, metal A only, metal B only, and bilayer of both metals each with nominal thickness of 1 nm. Raman mapping measurements during/after heating were conducted on a home-built system with 532 nm laser excitation and a 20× objective with numerical aperture of 0.45, under which the sample was mounted in a heating stage equipped with gas flow control and optical access through a quartz window. Raman spectra were acquired at discrete temperatures upon heating and cooling in various atmospheres (vacuum, Ar/$H_2$, and $H_2O$). Automated mapping over predefined regions at programmed temperatures was enabled by automated XY-stage control based on fiducial alignment patterns, maintaining spatial consistency across the entire heating process in a manner similar to the ref. [50].

*4.3.Kinetic Monte Carlo simulations on edge dynamics of nanotubes and graphene.*

Kinetic Monte Carlo (kMC) simulations were performed to model the edge dynamics during nanotube and graphene growth or etching [38,41]. For SWCNTs, the simulation assumes $C_2$ addition at anti-AC sites, with the probability proportional to the product of the carbon atom count in the catalyst and the number of anti-AC sites at the nanotube edge. The reverse reaction (etching) rate was proportional only to the AC site count. Reaction probabilities were determined by an Arrhenius-type expression with activation energies (taken as 1 eV) incorporating an interfacial energy difference ($E_Z-E_A$) between ZZ and AC edge atoms. Adsorption rates of carbon on catalyst are proportional to effective carbon concentration $P_C$ in gas phase, while desorption rates depend both on the carbon atom count $N$ in catalyst and effective concentration of etchants $P_E$, emulating the experimental picture in ref. [20]. Graphene simulations used a similar framework but incorporated additional pathways such as $C_3$ addition at ZZ edges forming new hexagons and ZZ atom removal generating notched edge structures, along with their respective reverse reactions. Energy barriers were selected based on literature-reported values [41]. Temperature used in kMC simulation was 1100 K for all the displayed data.

**Author Contributions**

**K. Otsuka:** Conceptualization, Funding acquisition, Methodology, Investigation, Software, Formal analysis, Visualization, Writing – original draft. **S. Maruyama:** Funding acquisition, Writing – review & editing.

*Corresponding author. Department of Mechanical Engineering, The University of Tokyo, Tokyo 113-8656, Japan.

*Email address*: otsuka@photon.t.u-tokyo.ac.jp (K. Otsuka)

**Acknowledgements**

A part of this work was supported by JSPS (KAKENHI JP22H01411, JP23H05443, JP21KK0087), JST (CREST JPMJCR20B5), the Ministry of Education, Culture, Sports, Science and Technology (MEXT), Japan, and Murata Science and Education Foundation. A part of this work was conducted at Takeda Sentanchi Supercleanroom, The University of Tokyo, supported by "Advanced Research Infrastructure for Materials and Nanotechnology in Japan (ARIM)" of MEXT (Proposal Number JPMXP1224UT1167).

**Supplementary data**

**Supplementary data for**

# Catalyst-mediated etching of carbon nanotubes exhibiting electronic-structure insensitivity and reciprocal kinetics with growth

Keigo Otsuka [a,*], Shigeo Maruyama [a,b,c]

[a] *Department of Mechanical Engineering, The University of Tokyo, Tokyo, 113-8656, Japan*

[b] *School of Mechanical Engineering, Zhejiang University, Hangzhou, 310027, China*

[c] *Institutes of Innovation for Future Society, Nagoya University, Nagoya 464-8601, Japan*

————

* email: otsuka@photon.t.u-tokyo.ac.jp

**Supplementary data**

**Materials and Methods**

*Kinetic model for growth and etching.* To quantitatively interpret the observed growth and etching rates, the kinetic model we previously proposed in ref. [1] is used with slight modifications in in this study. In this model, the net growth rate of a SWCNT $\gamma$ is governed by the balance between the carbon adsorption (supply) and carbon desorption (removal) on the catalyst. For simplicity, here we define all rates in terms of carbon atom flux (number of C atoms per unit time), rather than nanotube length basis. In the steady state, the growth (etching) rate is given by:

$$\gamma = \gamma_{ad} - \gamma_{de},$$

where $\gamma_{ad}$ is the adsorption rate of carbon atoms onto the catalyst from the feedstock gas, and $\gamma_{de}$ is the desoption rate due to etching agents. Both terms are assumed to follow first-order kinetics with respect to the relevant gas concentrations:

$$\gamma_{ad} = Ak_{ad}P_C,$$

$$\gamma_{de} = Ak_{de}P_E N,$$

where $P_C$ is the partial pressure of the carbon-containing feedstock (e.g., ethanol), $P_E$ is the partial pressure of etchants (e.g., water vapor), and $k_{ad}$ and $k_{de}$ are the respective reaction rate constants. Also, $A$ is the surface area of the catalyst, and $N$ is the number of carbon atoms adsorbed on the catalyst.

Under steady-state growth conditions with only ethanol as a carbon source, the balance shifts toward positive $\gamma$, whereas in the presence of oxidants alone (i.e., no ethanol), $\gamma$ becomes negative and corresponds to a net etching process. Explicit expression of the growth (etching) rate is as follows:

$$\gamma = \frac{Ak_g(k_{ad}P_C - k_{de}P_E N_{eq})}{k_g + Ak_{de}P_E}.$$

To quantify the intrinsic removal ability of water, we analyze the experimental result shown in Fig. S3b, where CNT growth rate suppression by water vapor was measured. By fitting the observed net growth rate $\gamma$ as a function of $P_E$ with $P_C$ held constant, the net desorption rate $\gamma_{de}$ becomes:

$$\gamma_{de} = A\big(k_{de}^{et}P_E^{et} + k_{de}^{w}P_E^{w}\big)N,$$

where $k_{de}^{w}$ and $P_E^{w}$ represent the kinetic constant for carbon removal and the effective pressure of etching agent derived from water, respectively. With $\eta \equiv k_{de}^{w}/k_{de}^{et}$ as a fitting parameter, we extract

**Supplementary data**

$\eta=1.45$, which represents the relative carbon removal ability of water vapor with respect to ethanol under equivalent partial pressure.

Using this ratio, we estimate the net etching rate in a condition without ethanol ($P_C=0$), as shown in Figure 2c. Specifically, we use the relationships:

$$\gamma = -\gamma_{de} = -A\eta k_{de}^{et} P_E^w N,$$

$$\gamma = k_g(N - N_{eq}),$$

where the average values are given in ref. [1] in the case of ethanol at the same temperature (800°C), except for $\eta$.

In addition, we introduce the following dimensionless sensitivity parameters to evaluate the influence of each reaction step:

$$S_i = \frac{\partial \ln \gamma}{\partial \ln k_i}$$

where $i$ can be replaced with ad, de, and g for each reaction. These parameters allow us to assess which process—adsorption, desorption, carbon precipitation—is rate-limiting under a given set of conditions. For instance, at low etchant pressures, $S_g$ tends to be small, indicating the rate-limiting step is adsoption and desorption of carbon, rather than the carbon exchange at the tube-catalyst junction. This is consistent to the weak correlation between growth rates and etching rates compared for the same tubes (Fig. 2b). Conversely, high water vapor concentrations lead to large $S_g$, indicating an enhanced influence of $k_g$, i.e., chiral-angle depedence, which mey help explain the chiral-angle preference of catalytic etching (Fig. S8g).

*Automated Raman mapping measurement during etching.* To investigate the temperature-dependent etching behavior of SWCNTs under controlled gas environments, a custom-built in-situ Raman spectroscopy system was developed. This system integrates a confocal Raman microscope with a optical cell (Linkam, 10042D) and gas flow control unitThe setup enables precise spectroscopic mapping of individual nanotubes or SWCNT networks during thermal annealing in various atmospheres such as vacuum, argon/hydrogen, and water vapor.

**Supplementary data**

The Raman system uses a 532 nm continuous-wave laser (Novanta, gem532), focused via a 20× objective lens (NA=0.45), which also collects the backscattered Raman signal. A long-pass edge filter removes Rayleigh scattering, and the filtered signal is sent to a spectrometer (Teledyne Princeton Instruments, HRS-300) equipped with a high-sensitivity CCD dector (Teledyne Princeton Instruments , PIXIS100BRX). All components are aligned in a backscattering configuration for optimal collection efficiency. Samples are mounted inside the sealed optical cell on a motorized XYZ stage, allowing for stable operation during temperature-controlled measurements in vacuum (1–3 Pa), Ar/$H_2$ (3% $H_2$), or water vapor (~100 Pa) atmospheres. For ensemble measurements, in which spatially averaged spectral features of CNT networks are of interest, a cylindrical lens is inserted into the beam path to elongate the laser spot into a vertical line. This reduces the excitation power density and enhances averaging over a broader sample area.

The system is fully automated via LabVIEW and performs spatially consistent Raman measurements throughout extended heating cycles. Cross-shaped trenches fabricated by dry etching are used as fiducial alignment markers on the silicon substrate. These enable automatic realignment of laser focus and position during heating and cooling cycles, compensating for thermal drift and mechanical shifts. This method is based on a previously reported system for photoluminescence measurements of individual SWCNTs [2].

A representative experiment shown in Figs. S6-8 illustrates the system's capability; a single substrate was pre-patterned into four distinct regions—no catalyst, Fe only, Pd only, and Fe/Pd dual layers, followed by the transfer of SWCNT networks. Raman mapping mesurements were sequentially perforned from each region at various temperatures up to 900°C and after cooling to the room tempratrure, with alignment recalibrated at each step. The entire procedure, including heating, cooling, gas switching, and spatial mapping, was conducted automatically over approximately one day. This allowed direct comparison of the etching dynamics between regions under identical thermal histories and atmospheric conditions, revealing subtle differences in reactivity arising from catalyst identity.

**Supplementary data**

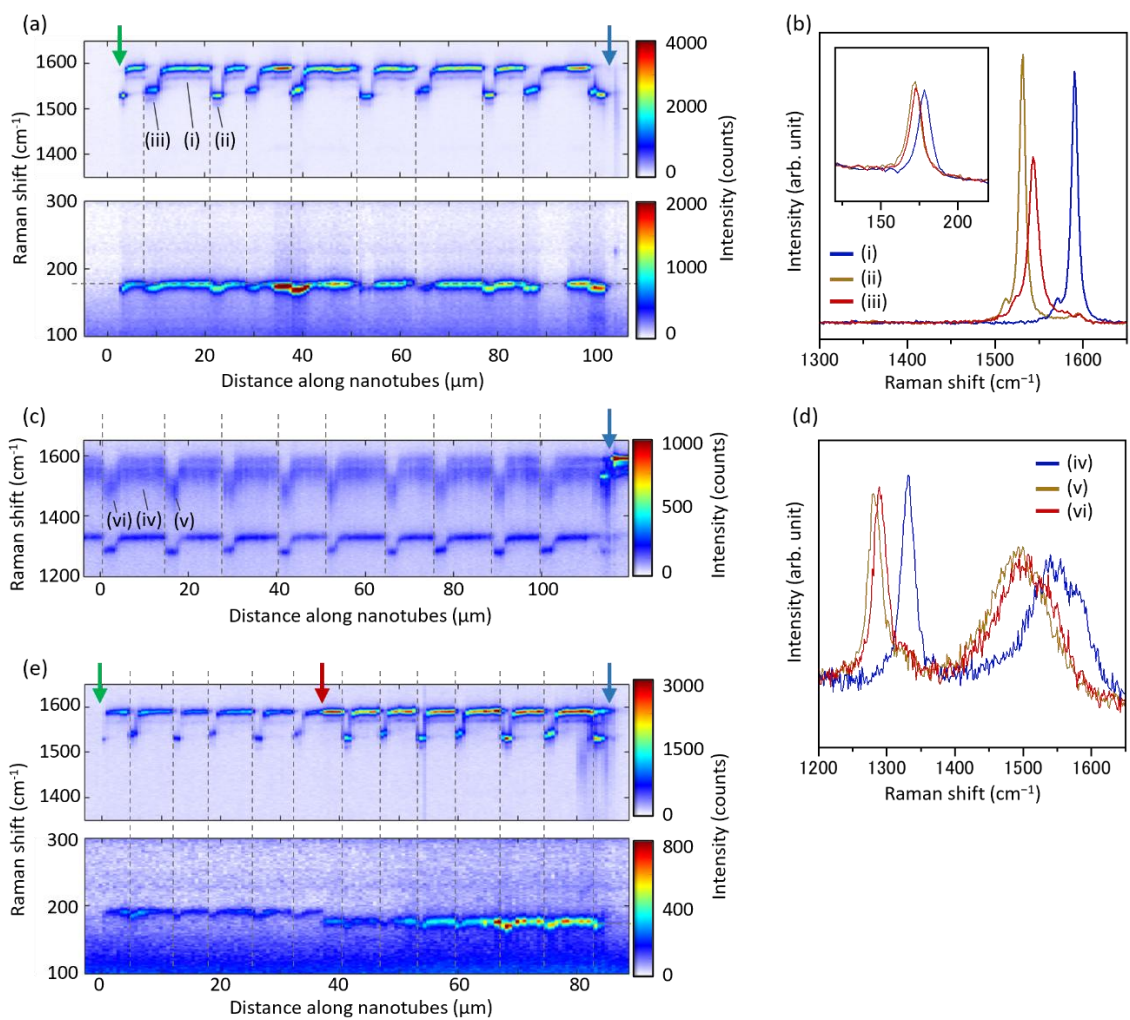

**Fig. S1 | Raman intensity maps for representative SWCNTs.** (a) Raman intensity is plotted as a color contour along the tube axis for a s-SWCNT. Green and blue arrows denote the tip and root of the nanotube, respectively. This result demonstrates the structural integrity of SWCNTs after multiple cycles of growth and etching. Transition of Raman spectra indicated by vertical dashed lines is sharp because no mixing occurs between $^{12}$C and $^{13}$C ethanol, separated by the etching or idling stages without carbon source. This is in good contrast to the other isotopic transition within the growth stage. (b) Raman spectra of the s-SWCNT at the position indicated by (i-iii) in (a). Isotope labels (ii) and (iii) grew from 100% and 80% $^{13}$C ethanol, respectively. (c,d) Similar Raman intensity map and spectra for a metallic CNT. (e) Raman intensity map for a rare s-SWCNT, whose chirality was changed during the growth at the position indicated by the red arrow. Switching among growth, idling, and etching occurs near the isotope label positions, suggesting that this chirality change is not directly caused by changes in the gas environment.

**Supplementary data**

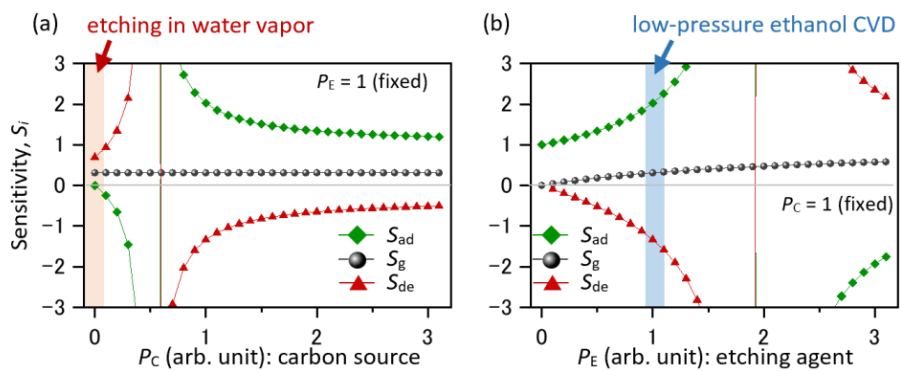

**Fig. S2 | Analysis of sensitivity of growth (etching) rate to each kinetic constant.** (a,b) Analytical sensitivity of growth rate γ as a function of effective pressure of carbon sources $P_C$ (a), and that of etching agents $P_E$ (b), which is obtained using the kinetic model in ref. [1]. Sensitivity $S_i$ ($i$=ad, de, and g) is defined as $S_i = \frac{\partial \ln \gamma}{\partial \ln k_i}$, where $k_{ad}$, $k_{de}$, and $k_g$ represent the kinetic constants for carbon adsorption on the catalyst, carbon desorption from the catalyst, and precipitation of carbon as a nanotube from the catalyst, respectively. Here, the growth condition with $P_C$=1 and $P_E$=1 roughly corresponds to our ethanol CVD process. The etching stage without ethanol supply can be considered as the condition with $P_C$=0 and $P_E$~1. When the pressure of etching agent becomes higher (large $P_E$), the influence of $k_g$ in net etching rates, i.e., chiral-angle dependence, can exceed that of $k_{de}$ from catalysts because the rate-limiting step shifts to catalyst-tube interface (see Fig. S8).

**Supplementary data**

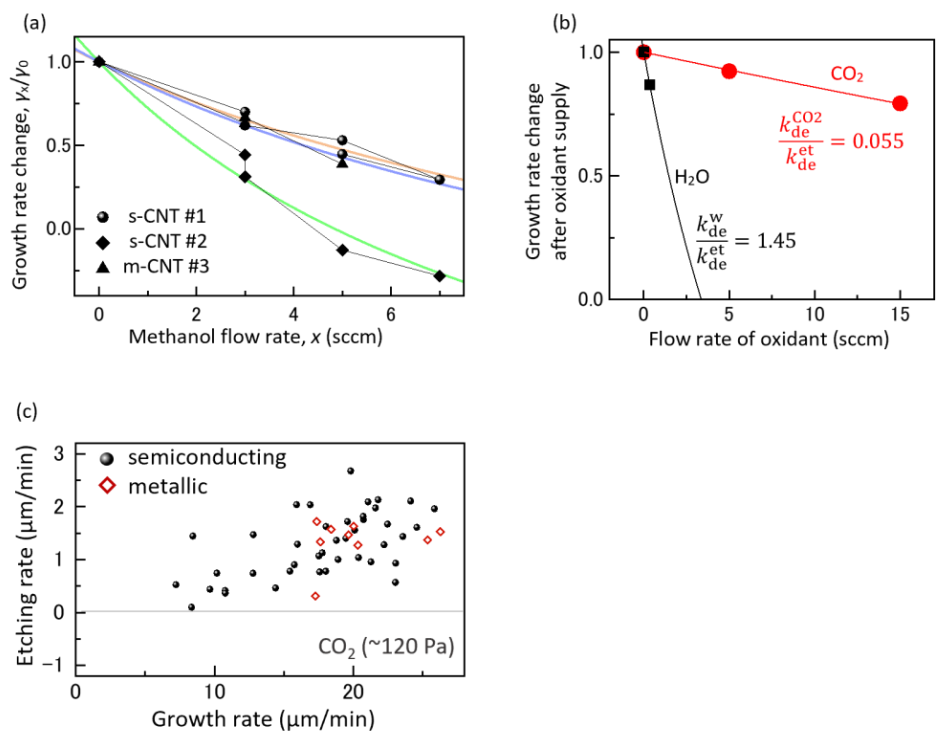

**Fig. S3 | Relative change of growth rate upon the addition of etching agents.** (a) Relative growth rate change as a function of methanol flow rate. Examples of three different isolated SWCNTs are shown. Solid lines represent the fitting based on the kinetic model with $\eta = k_{de}^{me}/k_{de}^{et}$ as the only fitting parameter, illustrating a smooth transition from growth to etching by increasing the methanol flow rate. (b) Relative growth rate change upon the supply of two types of pure oxidants: $H_2O$ and $CO_2$. Unlike in the panel (a), the data were averaged over many SWCNTs. Fitting with the kinetic model yields the relative etching efficiency of the oxidants in the form of $\eta$. (c) Growth rates versus etching rates for individual SWCNTs when $CO_2$ (~120 Pa) was used as an oxidant to induce etching.

**Supplementary data**

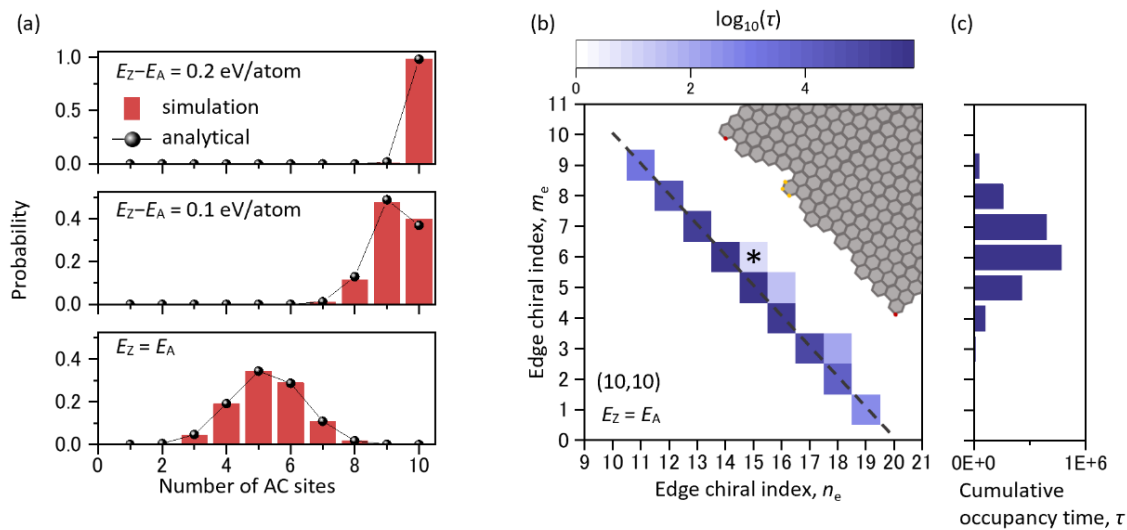

**Fig. S4 | Kinetic Monte Carlo simulation of edge configurations.** (a) Probability distribution of the number of AC sites at the CNT edge as a function of the energy difference between zigzag atoms and armchair atoms ($E_Z$–$E_A$). Filled bars and circles represent the probability obtained from kinetic Monte Carlo (kMC) simulations (filled bars) and the analytical degeneracy $g_i$ [3] weighted with a Boltzmann factor (circles). (b) Two-dimensional histogram of edge configurations for (10,10) nanotube obtained from the same kMC framework used for graphene growth. The edge chiral indices ($n_e$,$m_e$), defined in ref. [4], are used to classify the edge structures. Inset: a representative edge structure at the star-marked configuration, with yellow-highlighted atoms that do not conform to conventional armchair or zigzag classifications. Such atoms arise from $C_1$ removal or $C_3$ addition. (c) Probability distribution of $m_e$ (number of AC sites at the edge) when $C_1$ and $C_3$ removal/addition is allowed. The distribution closely resembles the distribution in Fig. 3d of the main text, confirming that the emergence of symmetric edge configurations in nanotubes is not merely a consequence of restricting the reaction pathways to $C_2$ addition or removal.

**Supplementary data**

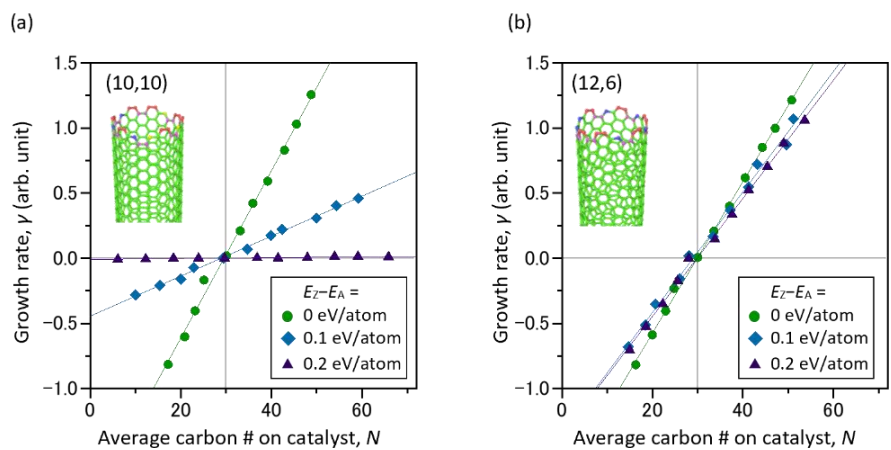

**Fig. S5 | Kinetic Monte Carlo simulation of growth rates and etching rates.** kMC simulations were performed to evaluate the relationship between the number of carbon atoms *N* stored in the catalyst (30 atoms at equilibrium) and the net nanotube growth rate. (a) Simulated growth rate of a (10,10) nanotube as a function of catalyst carbon count under various edge energy asymmetries ($E_Z$–$E_A$), where $E_Z$ and $E_A$ represent the energy of armchair and zigzag edge atoms, respectively. The carbon source pressure ($P_C$) was varied while the etchant pressure ($P_E$) was fixed. (b) Corresponding net growth (or etching) rates for a (12,6) nanotube, using the same parameter set. In both chiral indices, growth and etching rates show linear dependence on catalyst carbon content, where the slope of fitted line corresponds to the kinetic constant $k_g$. These results validate the kinetic symmetry between growth and etching and support the underlying model assumption that both processes are governed by deviations from carbon equilibrium within the catalyst.

**Supplementary data**

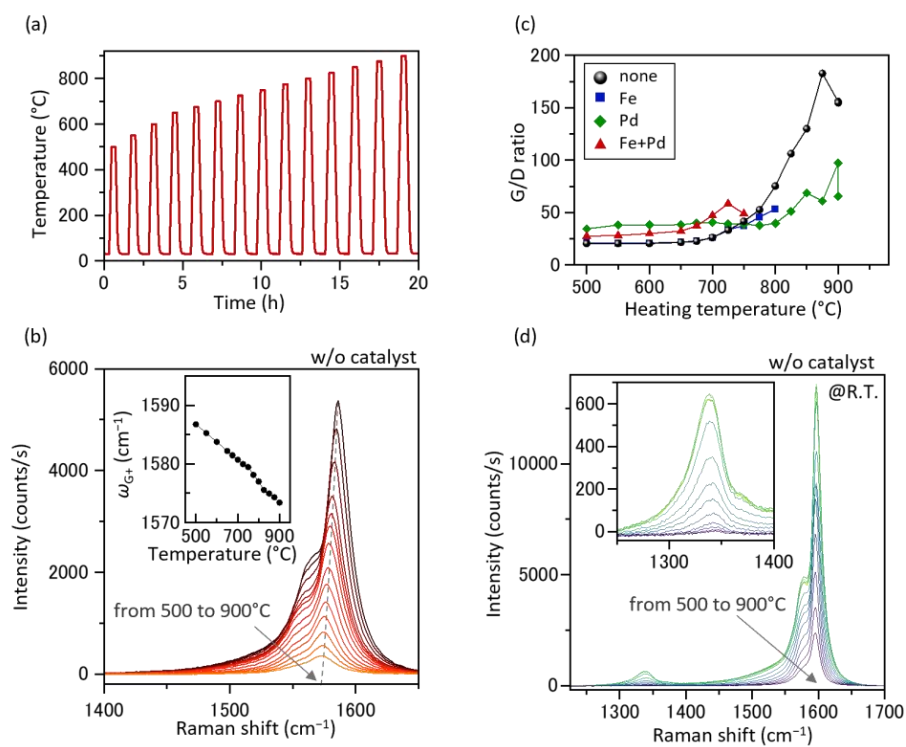

**Fig. S6 | In-situ Raman measurement during/after catalytic and non-catalytic etching of SWCNT networks in water vapor.** (a) Temperature history of the heating cell. Fe and Pd were used as catalysts. The heating cell was filled only with ~100-Pa water vapor. (b) G-band spectra of SWCNTs measured at high temperatures. Inset: Peak frequency of G$^+$ mode at different temperatures. (c) G/D ratios of four different catalytic conditions: no catalysts, Fe, Pd, Fe and Pd. The nominal thickness of each metal was 1 nm. All the Raman spectra were measured at room temperature. Spectra with G-mode intensity below 10,000 counts/s are excluded due to high uncertainty. Initial G/D ratio after the polymer-mediated transfer process was ~7. (d) Raman spectra measured after etching processes at different heating temperatures up to 900°C. Inset: Zoomed up D-mode spectra. All the spectra were obtained with excitation laser wavelength of 532 nm.

**Supplementary data**

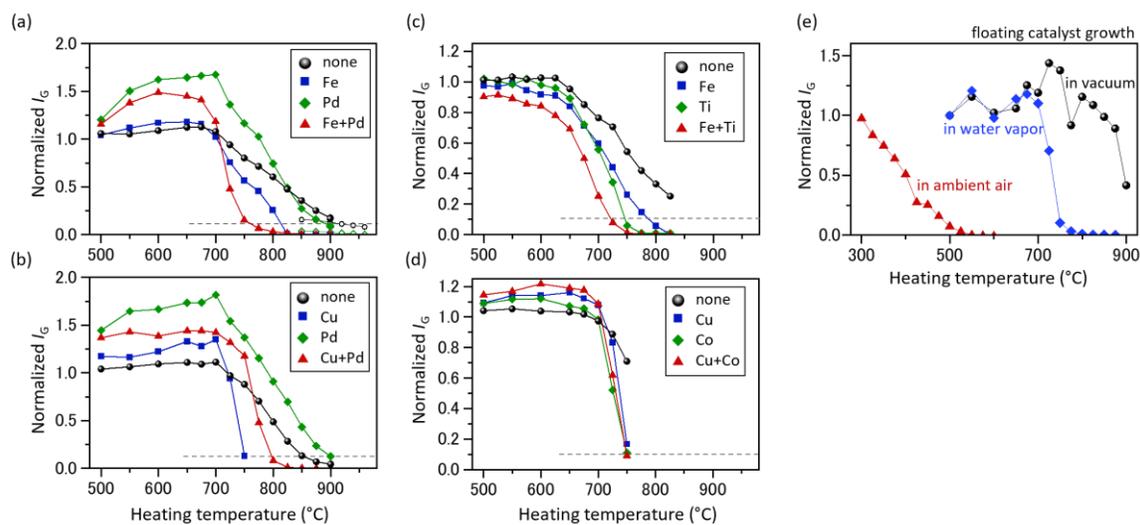

**Fig. S7 | Evolution of relative G-mode intensity $I_G$ after etching process.** (a-d) $I_G$ normalized to those measured after annealing in vacuum at 600°C. Catalysts deposited below SWCNT networks were Fe, Pd, and Fe/Pd (a), Cu, Pd, and Cu/Pd (b), Fe, Ti, and Fe/Ti (c), Cu, Co, and Cu/Co (d). All the etching was enhanced by water vapor (~100 Pa). Note that open circles and open diamonds in panel (a) represent $I_G$ evolution in vacuum (1–2 Pa). (e) Similar trace of normalized $I_G$ in vacuum (residual air of ~1 Pa, black circles), in water vapor (~100 Pa, blue diamonds), and in ambient air (red triangles). Only in panel (e), CNTs grown from floating catalyst CVD were used, in which ferrocene-derived Fe nanoparticles were deposited on the CNT film. All the Raman measurements were performed with excitation laser wavelength of 532 nm.

**Supplementary data**

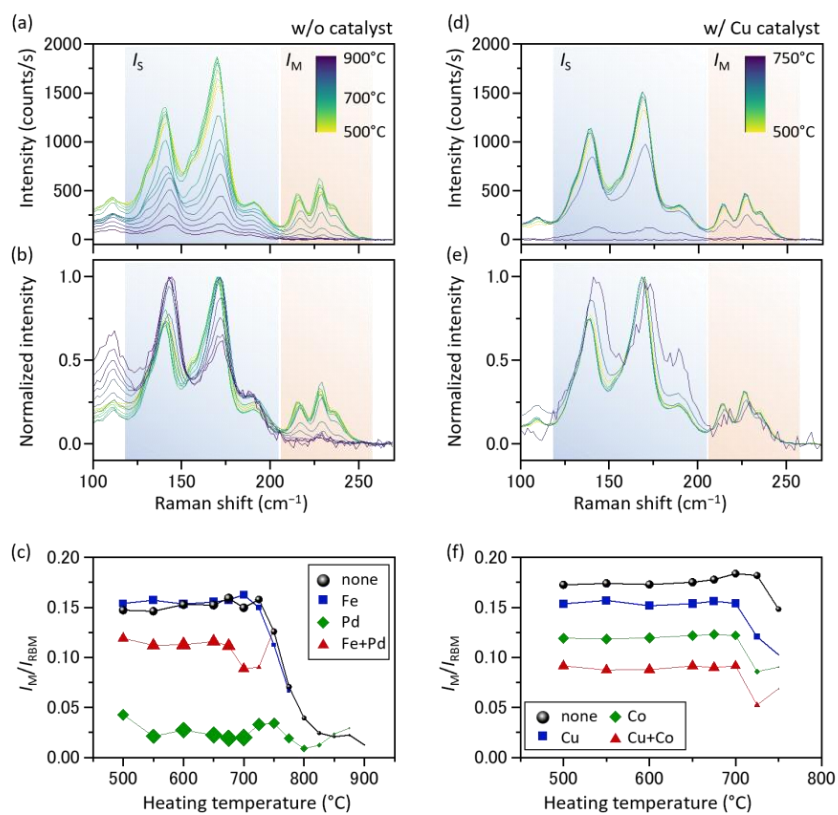

**Fig. S8 | Evolution of RBM spectra after successive etching process.** (a) RBM spectra measured after non-catalytic etching in water vapor (~100 Pa). (b) Normalized RBM spectra, indicating a relative increase in large diameter SWCNTs. (c) Ratio of Raman intensity integrated over 206–270 cm$^{-1}$ to the intensity integrated over the entire RBM region. Substrate-derived background signals were subtracted from all spectra. The width of each mark is proportional to the normalized $I_G$ shown in Fig. S7. Although the $I_M/I_{RBM}$ ratios decreased in the Fe catalyst region, the extent of reduction closely matches that observed under non-catalytic conditions. This agreement, despite the decrease in G-mode intensity at lower temperatures (Fig. S7) in the catalyst region, suggests that the observed spectral changes are attributed to non-catalytic etching simultaneously taking place under the high water vapor pressure. (d–f) Similar RBM analysis for the SWCNTs transferred on Cu catalyst, showing a stark contrast to the case without catalyst. This likely reflects the high efficiency of Cu-mediated etching that occurred below 750°C.

**Supplementary data**